\documentclass{emulateapj}
\usepackage{psfig}
\usepackage{graphicx}
\usepackage{graphics}

\newcommand{\ssst}{\scriptscriptstyle}
\newcommand{\E}[1]{\times 10^{#1}}
\newcommand{\RA}[3]{{#1}^{{\rm h}}{#2}^{{\rm m}}{#3}^{{\rm s}}}
\newcommand{\Dec}[3]{{#1}^{\circ}{#2}'{#3}''}

      \newcommand{\ps}{\,{\rm s}^{-1}}
    \newcommand{\Msun}{M_{\odot}}
\newcommand{\cm}{\,{\rm cm}}    \newcommand{\km}{\,{\rm km}}
\newcommand{\parsec}{\,{\rm pc}}\newcommand{\kpc}{\,{\rm kpc}}
\newcommand{\keV}{\,{\rm keV}}

        \newcommand{\NH}{N_{\ssst\rm H}}
        
\newcommand{\nHH}{n({\rm H}_{2})} \newcommand{\NHH}{N({\rm H}_{2})}
 \newcommand{\ASCA}{{\sl ASCA}}
\newcommand{\Chandra}{{\sl Chandra}}
\newcommand{\XMM}{{\it XMM-Newton}}
\newcommand{\Spitzer}{{\sl Spitzer}}
\newcommand{\du}{d_{10.3}}
\newcommand{\twCO}{$^{12}$CO}  \newcommand{\thCO}{$^{13}$CO}

\begin{document}

\title
{Cavity of Molecular Gas Associated with Supernova Remnant 3C~397}

\author
{Bing Jiang\altaffilmark{1},
Yang Chen\altaffilmark{1,2,3},
Junzhi Wang\altaffilmark{1,2},
Yang Su\altaffilmark{4},
Xin Zhou\altaffilmark{1},
Samar Safi-Harb\altaffilmark{5}, and
Tracey DeLaney\altaffilmark{6}}
\altaffiltext{1}{Department of Astronomy, Nanjing University, Nanjing 210093,
China}
\altaffiltext{2}{Key Laboratory of Modern Astronomy and Astrophysics
(Nanjing University), Ministry of Education, China}
\altaffiltext{3}{Corresponding author}
\altaffiltext{4}{Purple Mountain Observatory, Chinese Academy of Sciences,
Nanjing 210008, China}
\altaffiltext{5}{Department of Physics and Astronomy, University of Manitoba,
Winnipeg, MB R3T 2N2, Canada}
\altaffiltext{6}{Kavli Institute for Astrophysics and Space Research, Massachusetts
Institute of Technology, 70 Vassar Street, Cambridge MA 02139, USA}





\begin{abstract}
3C~397 is a radio and X-ray bright Galactic supernova remnant
(SNR) with an unusual rectangular morphology.
Our CO observation obtained with the Purple
Mountain Observatory at Delingha reveals that the remnant is well confined
in a cavity of molecular gas,
and embedded at the edge of a molecular cloud (MC) at
the local standard of rest systemic velocity of $\sim32\km\ps$.
The cloud has a column density gradient increasing from southeast
to northwest, perpendicular to the Galactic plane, in agreement
with the elongation direction of the remnant.
This systemic velocity places the cloud
and SNR 3C~397 at a kinematic distance of $\sim10.3\kpc$.
The derived mean molecular density ($\sim10$--$30\cm^{-3}$)
explains the high volume emission measure of the X-ray emitting gas.
A \twCO\ line broadening of the $\sim32\km\ps$ component is detected
at the westmost boundary of the remnant, which provides direct evidence
of the SNR-MC interaction and suggests multi-component gas there with
dense ($\sim 10^4 \cm^{-3}$) molecular clumps.
We confirm the previous detection of
a MC at $\sim38\km\ps$ to the west and south of the SNR and
argue, based on HI self-absorption,
that the cloud is located in the foreground of the remnant.

A list of 64 Galactic SNRs presently known and suggested to be in
physical contact with environmental MCs is appended in this paper.
\end{abstract}


\section{INTRODUCTION}
Core-collapse supernovae are often not far away from their natal
giant molecular clouds (MCs).
About half of the Galactic supernova remnants
(SNRs) are expected to be in physical contact with MCs
(Reynoso \& Mangum 2001),
but so far only less than 40 SNR-MC associations have been confirmed (see Appendix).
These associations are mostly established by the detection
of OH 1720 MHz masers (e.g., Frail et al.\ 1996; Green et al.\ 1997).
For the rest few cases, the convincing evidence comes from the sub-mm/mm
observations of molecular lines and infrared observations in recent
decades, including molecular line broadening, high line ratio (e.g.,
\twCO\ $J=2$--$1/J=1$--0), morphological correspondence of molecular
emission, etc.
Many SNR-MC associations may have not yet been revealed mainly because the
OH maser emission is below the detection thresholds
(Hewitt \& Yusef-Zadeh 2009).
Even for most of the SNRs 
whose interaction with MCs are confirmed,
the detailed distribution of environmental molecular gas,
which can shed light on
the SNRs' dynamical evolution and physical properties,
is poorly known.
In this regard, line emission of CO and its isotopes plays an important
role in the study of the SNR-MC association.

3C~397 (G41.1$-$0.3) is a
radio and X-ray bright Galactic SNR
with a peculiar rectangular morphology.
It is elongated along the southeast (SE)-northwest (NW) direction,
perpendicular to the Galactic plane \citep{b1}. 
In X-rays, a ``hot'' spot was detected near the geometric center
(Chen et al.\ 1999; Dyer \& Reynolds 1999; Safi-Harb et al.\ 2000),
 suggestive of a compact object associated with the SNR;
however,  using the \ASCA\ and \Chandra\ observations,
no pulsed signal and no hard, non-thermal emission were detected from it
as would be expected from a neutron star or a pulsar wind nebula
(Chen et al.\ 1999; Safi-Harb et al.\ 2000; Safi-Harb et al.\ 2005,
hereafter S05).
Based on the irregular morphology seen in the radio and X-rays,
the sharp western boundary,
and the higher density in the western region of the SNR,
S05 suggested that SNR 3C~397 is
encountering or interacting with a MC to the west.
Using millimeter data to study the environs of 3C~397,
they found evidence for a MC at a
local standard of rest (LSR) velocity of $V_{\rm LSR}=35$--$40\km\ps$
to the west and south of the remnant,
and suggested a possible association between the cloud and the SNR.
The non-detection of OH masers toward the SNR and the small
$^{12}$CO ($J$=2--1)/($J$=1--0) ratio
led S05 to conclude that
either the MC is in the close vicinity of the SNR
but has not yet been overrun by the shock wave, or that
if the SNR-MC interaction is occurring, it must
be taking place along the line of sight.

Motivated by the above findings and
aiming to investigate the detailed distribution
of the molecular gas environing SNR 3C~397,
we performed a new observation in \twCO\ ($J$=1--0)
and \thCO\ ($J$=1--0) lines towards the remnant.
In this paper, we present our discovery
of a cavity of molecular gas associated with 3C~397,
and argue that the $38\km\ps$ cloud is likely in the
foreground of the SNR.

\section[]{OBSERVATIONS}

The observation was taken
with the 13.7~m millimeter-wavelength telescope of the Purple Mountain
Observatory at Delingha (hereafter PMOD), China,
 during November to December, 2008.
An SIS receiver was used to simultaneously observe the $^{12}$CO
($J$=1--0) line (at 115.271 GHz)
and $^{13}$CO ($J$=1--0) line (at 110.201 GHz), while
two acousto-optical spectrometers were used as the back end with
1024 channels. The corresponding spectral coverage was 145 MHz for
$^{12}$CO ($J$=1--0) and 43 MHz for $^{13}$CO ($J$=1--0).
We mapped a $4'\times 3'.5$ area centered at
($\RA{19}{07}{34}.0, \Dec{+07}{08}{00}.0$, J2000.0),
which covers 3C~397 via raster-scan mapping with
a grid spacing of $30''$.
The half-power beamwidth of the telescope was $60''$ and
the main-beam efficiency was 52\% during the observation epoch.
The typical system temperature was around 180--260 K.
The observed LSR velocity ranges
were $-150$ to +230$\km\ps$
for $^{12}$CO and $-19$ to +99$\km\ps$ for $^{13}$CO. The velocity
resolution was $0.5\km\ps$ for $^{12}$CO and $0.2\km\ps$ for $^{13}$CO.
After performing elevation calibration and baseline subtraction with a
low-order polynomial fit, we got an average
rms noise of $\sim 0.3$ K for both $^{12}$CO and $^{13}$CO data.
All the CO data were reduced using the GILDAS/CLASS
package \footnote{http://www.iram.fr/IRAMFR/GILDAS}.
The quality of the new CO observation is generally similar to that of
the SEST and GPS data used in S05, except that the grid points for
\twCO\ ($J$=1--0) are set denser than those in the SEST data.

\begin{figure}
\vspace{-32mm}
\hspace{3mm}
\psfig{figure=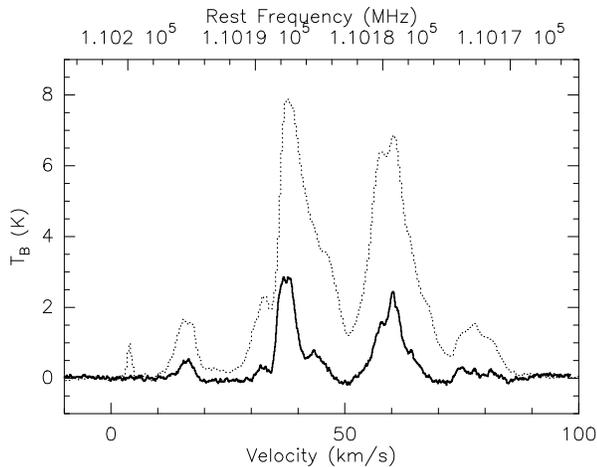,height=0.55\textwidth,angle=270, clip=}
\caption{The average CO spectra over the FOV. The dashed line is
$^{12}$CO ($J$=1--0) and the solid line is $^{13}$CO ($J$=1--0)
(multiplied by a factor of 2). } \label{f:spec}
\end{figure}

In order to correlate the distribution of molecular material
in the environs of 3C~397 with the morphology of the SNR,
we used the archival \Chandra\ X-ray (ObsID: 1042, PI: S.~S.~Holt)
and \Spitzer\ $24\mu$m mid-infrared (IR) (PID: 3483, PI: J.~Rho)
data.
The VLA 1.4~GHz radio continuum emission (L~band) data were adopted from
\citet{b1}, 
and the HI line emission data were obtained from the VLA Galactic Plane
Survey (VGPS) (Stil et al.\ 2006).

\section{RESULTS}
\subsection{The Molecular Gas Cavity at $V_{\rm LSR}\sim32\km\ps$}
\label{sec:gas32km}

\begin{figure*}
\vspace{-2mm}
\centerline{
\psfig{figure=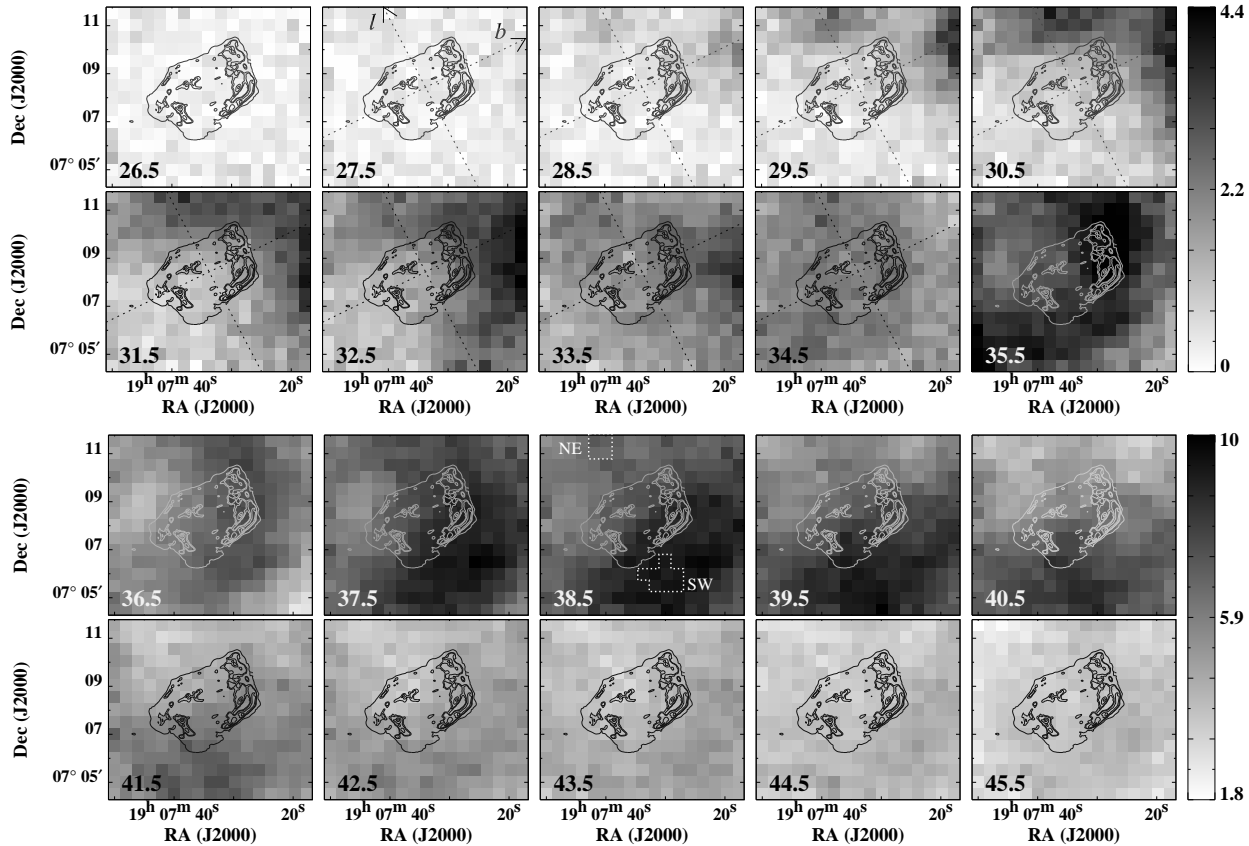,height=4.6in,angle=0,clip=} }
\caption{ $^{12}$CO ($J$=1--0) intensity maps integrated
each $1\km\ps$ in the velocity range of 26-46$\km\ps$, overlaid by
VLA 1.4 GHz radio continuum emission in contours with levels of 2,
9, 13.5, 18, 23, 32 mJy beam$^{-1}$. The central velocities are
marked in each image. The average rms noise of each map is 0.18
K$\km\ps$. The dashed boxes labeled with ``NE'' and ``SW'' denote
the regions from which the \thCO\ and HI spectra (see
Fig.~\ref{f:HISA}) were extracted. The dashed lines indicate lines
of Galactic latitude and longitude along which the column density
distribution $N(\rm H_2)$ of the $32\km\ps$ \twCO\ molecular
component was derived (see Fig.~\ref{f:column}). } \label{f:vmap}
\end{figure*}

Figure~\ref{f:spec} shows the \twCO\ ($J$=1--0) and \thCO\ ($J$=1--0)
spectra
averaged over the field of view (FOV).
Several velocity components are present in the velocity range
$V_{\rm LSR}=0$--$100\km\ps$, 
and no CO emission is detected outside of this range.
There are two prominent CO emission peaks around $38\km\ps$
and $55\km\ps$.
We made $^{12}$CO emission intensity channel maps with
$\sim1\km\ps$ velocity intervals.
No evidence is found for the positional correlation between the
55\,(50--$70)\km\ps$ CO component and SNR 3C~397.
The other prominent CO emission peak at $38\km\ps$ appears to form a
crescent strip that partially surrounds the western and southern borders of
the remnant as shown in the 35--$42\km\ps$ velocity interval of
Figure~\ref{f:vmap}.
This is consistent with the structure seen in
the \thCO\ images in the 35.4--$41.3\km\ps$ interval
found by S05.

\begin{figure}
\vspace{6mm}
\centerline{
\psfig{figure=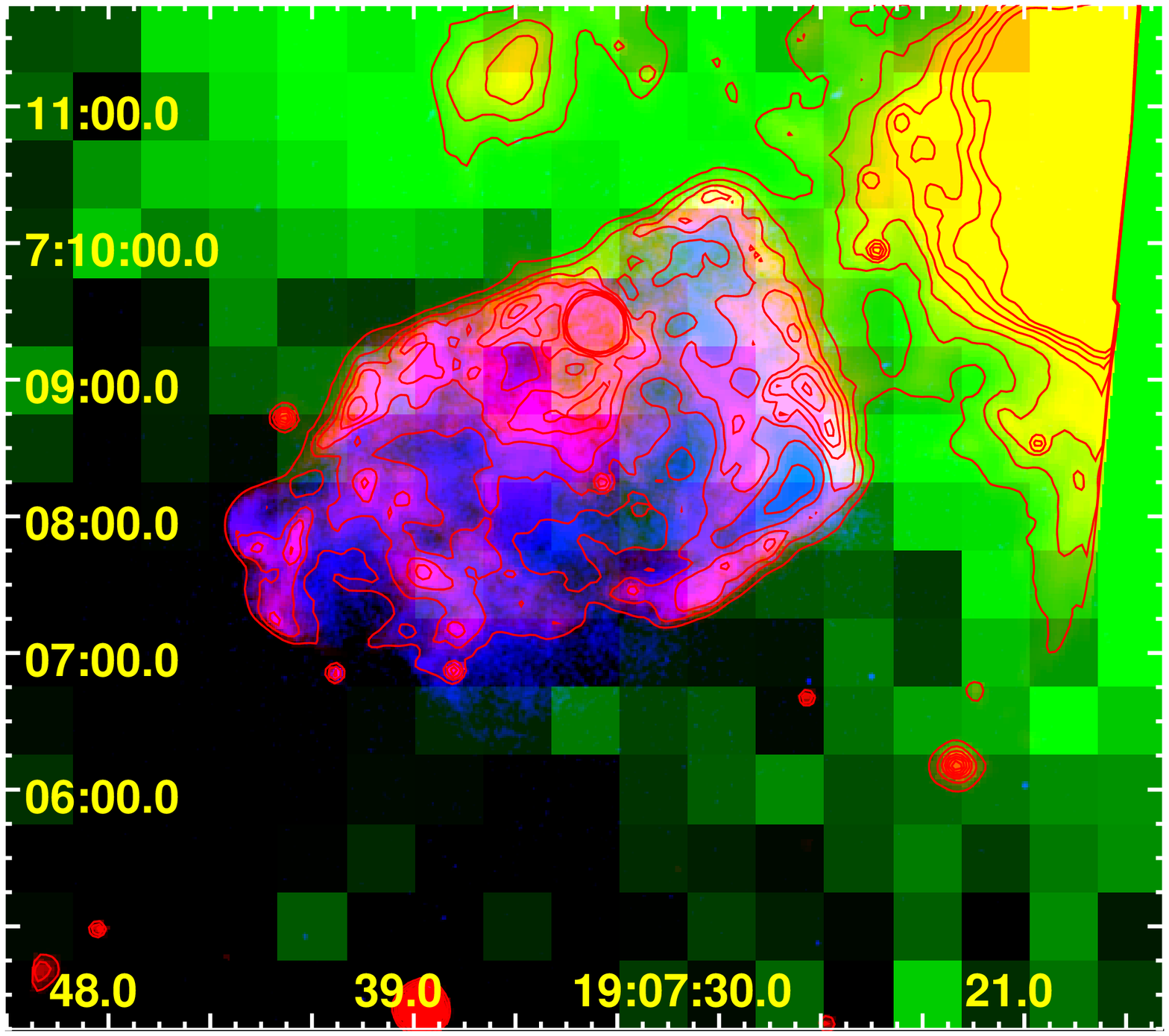,width=0.49\textwidth,angle=0, clip=}
}
\caption{
$^{12}$CO ($J$=1--0) intensity map in the velocity interval 27--$35\km\ps$
(above $6\sigma$)
coded in green, overlaid with the \Chandra\ X-ray image in blue, \Spitzer\
$24\mu$m mid-IR image in red (
with seven contours at logarithmic levels
between 68 and 120 MJy~sr$^{-1}$).
The bright yellow in the northwestern corner shows a part of
the foreground HII region G41.096$-$0.213 at $V_{\rm LSR}\sim59\km\ps$.
}
\label{f:multi}
\end{figure}

Interestingly, the intensity maps in the 27--$35\km\ps$ interval of
Figure~\ref{f:vmap} reveal a ``bay''-like cavity of molecular gas that
coincides with the remnant well.
The SNR appears surrounded by the molecular gas except in the SE.
As seen in the CO spectra (Fig.~\ref{f:spec}), the molecular gas
in this range corresponds to a molecular component peaked at $32\km\ps$.
Although not as prominent as the $38\km\ps$ component, the
$32\km\ps$ component
can be resolved in both the \twCO\ and \thCO\ spectra.
Because \thCO\ ($J$=1--0) is usually optically thin and indicative of
a high column density of H$_2$ molecules,
the presence of the \thCO\ peak ($\ga6\sigma$) at $\sim$ $32\km\ps$
implies that it is not a broadened part from the left wing of the
$38\km\ps$ peak, but a separate component.

Figure~\ref{f:multi} shows the integrated \twCO\ map in the velocity range
of 27--$35\km\ps$ overlaid with the \Chandra\ 0.5--8~keV X-ray image
and the \Spitzer\ 24$\mu$m mid-IR image of the SNR.
The mid-IR image displays a distinct SNR ``shell'' in the north, west,
and southwest,
and has a morphology strikingly similar to that in radio and X-rays.
In particular, it is worth noting the sharp western boundary seen at all
wavelengths.
The box-like shell appears to be well confined in the molecular gas
cavity, both of which seem open in the SE.
Furthermore, the flat northeastern boundary appears to
follow a sharp \twCO\ intensity interface.
This is consistent with a possible density enhancement in the north as
suggested by Anderson \& Rudnick (1993) for the inhibition of the SNR
expansion in this direction.

\begin{figure}
{\hfil
\psfig{figure=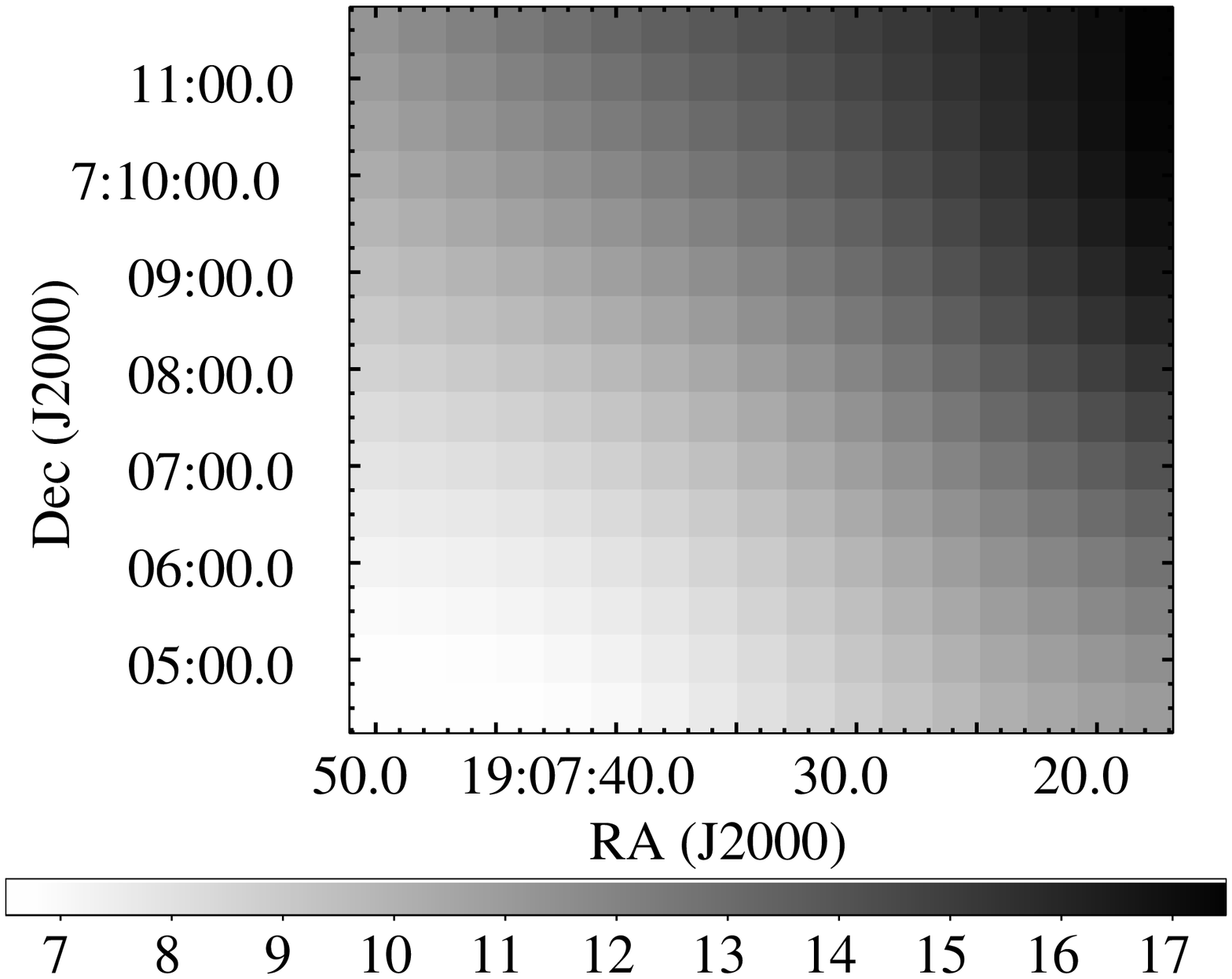,height=0.195\textwidth,angle=0, clip=}
\psfig{figure=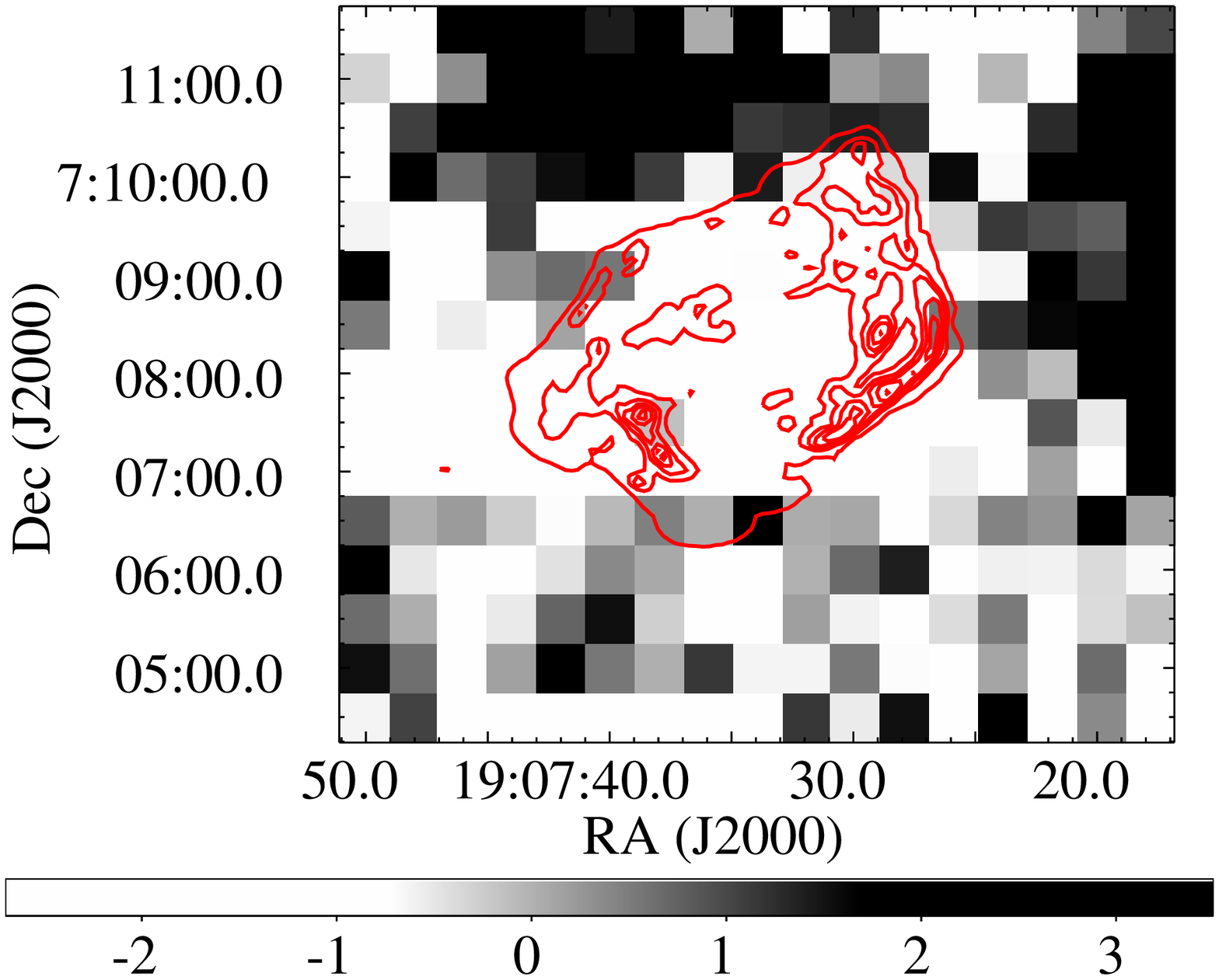,height=0.195\textwidth,angle=0, clip=}
\hfil}
\caption{
Left panel: the $^{12}$CO data intensity map in the same velocity interval
as in Fig.~\ref{f:multi},
smoothed to $2.5'$ to illustrate the large-scale confusion in this
direction.
Right panel: the same $^{12}$CO data as in Fig.~\ref{f:multi} after
unsharp masking (see text in \S~\ref{sec:gas32km}), overlaid by VLA 1.4 GHz contours
with the same levels as in Fig.~\ref{f:vmap}.
}
\label{f:mask}
\end{figure}

The 27--$35\km\ps$ \twCO\ intensity gradient from SE to NW
is clearly seen in Figures~\ref{f:vmap} and~\ref{f:multi}.
This is consistent with the large-scale density gradient
which was suggested based on the SNR brightening in radio
towards the Galactic plane (Anderson \& Rudnick 1993).
In order to remove the confusion by this broad structure at
low Galactic latitude ($-0.3^{\circ}$),
we applied the unsharp masking method as described in Landecker
et al.\ (1999) in which
we subtracted the smoothed \twCO\ intensity map
(Fig.~\ref{f:mask}, left panel)
from the original \twCO\ map.
In the resultant ``cleaned'' image (Fig.~\ref{f:mask}, right panel),
the remnant sits within a \twCO\ emission void again.
A similar result can also be obtained by subtracting
the smoothed and scaled $^{13}$CO 27--$35\km\ps$ intensity
map from the \twCO\ image.
The \twCO\ emission seems weak along the southwestern border of the SNR
in both Figures~\ref{f:multi}~\&~\ref{f:mask}.
This may be because the intensity is integrated only to $35\km\ps$
(due to the overlap of the left wing of the $38\km\ps$ component,
see Fig.~\ref{f:spec})
so that the contribution from the right wing of the $32\km\ps$
component is not taken into account.
Beyond $35\km\ps$,
the \twCO\ emission of the crescent strip (mentioned above) at
around $38\km\ps$
is strong in the southwest (see Fig.~\ref{f:vmap}) and
does not allow us to unambiguously distinguish the contribution there
from the right wing of the $32\km\ps$ component.

As described above, both the $38\km\ps$ and the $32\km\ps$
CO velocity components are shown to have a spatial correlation
with SNR 3C~397.
By comparison, the morphological correspondence of the former
is not as good as the latter because there is no molecular density
enhancement in the northeast around $38\km\ps$ to confine the remnant.

Applying the Milky Way's rotation curve of Clemens (1985)
together with $R_{0}=8.0 \kpc$ (Reid 1993) and $V_{0}=220\km\ps$,
the systemic LSR velocity of $32\km\ps$ corresponds to two candidate
kinematic distances, 1.7~kpc (near side) and 10.3~kpc (far side).
For $38\km\ps$ the kinematic distances are
2.1~kpc and 9.9~kpc.
For comparison, the tangent point in the direction to 3C~397
is at $\sim6.0\kpc$.
The discrepancy of the kinematic distances to the two molecular components
is at least 0.4 kpc, much larger than the typical size of a giant
MC complex ($\sim50$~pc; Cox 2000).
Thus it is almost impossible for SNR 3C~397 to simultaneously
interact with the two clouds.

\begin{figure}
\vspace{-5mm}
\centerline{
\psfig{figure=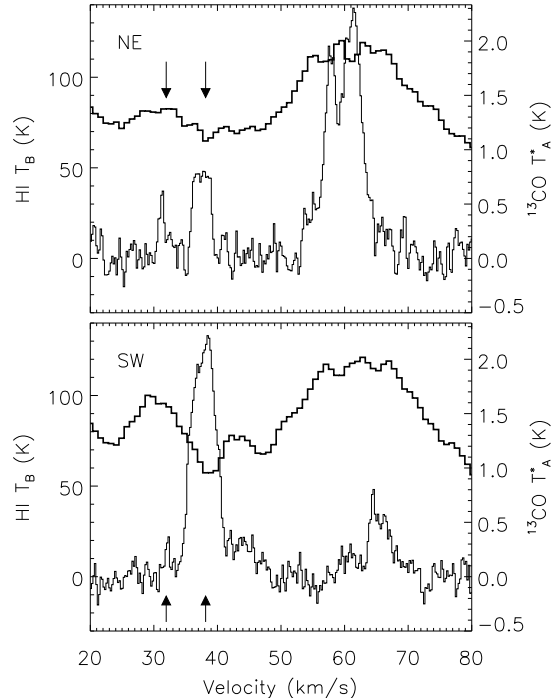,height=0.56\textwidth,angle=0, clip=}
}
\caption{
VGPS HI (thick line) and PMOD \thCO\ (thin line) spectra for the
NE (upper panel) and SW (lower panel) regions marked in Fig.~\ref{f:vmap}.
The arrows denote the locations of the LSR velocities of $32\km\ps$ and
$38\km\ps$.
}
\label{f:HISA}
\end{figure}

We can use the HI self-absorption (SA) measurement to help resolve
the near-far ambiguity in kinematic distances,
as suggested by Liszt et al.\ (1981) and Jackson et al.\ (2002).
The likelihood of detecting HI~SA favors the geometry
in which a cloud lies at the near kinematic distance
(see Fig.~\ref{f:spec} in
Roman-Duval et al.\ 2009 for this reasoning).
Figure~\ref{f:HISA} shows the VGPS HI and PMOD \thCO\ spectra of the
NE and SW regions near the SNR.
The HI~SA is clearly seen at $38\km\ps$ associated with the \thCO\
peak, which indicates that this cloud is most probably at the
near distance, 2.1~kpc.
On the contrary, little HI~SA is detected at $32\km\ps$.
It has been suggested that SNR 3C~397 is located beyond 7.5 kpc and
behind the tangent point based on the HI absorption
(Radhakrishnan et al.\ 1972; Caswell et al.\ 1975).
Therefore, the $38\km\ps$ velocity component is likely to be a
foreground MC towards the SNR.

\begin{figure*}
\centerline{
\psfig{figure=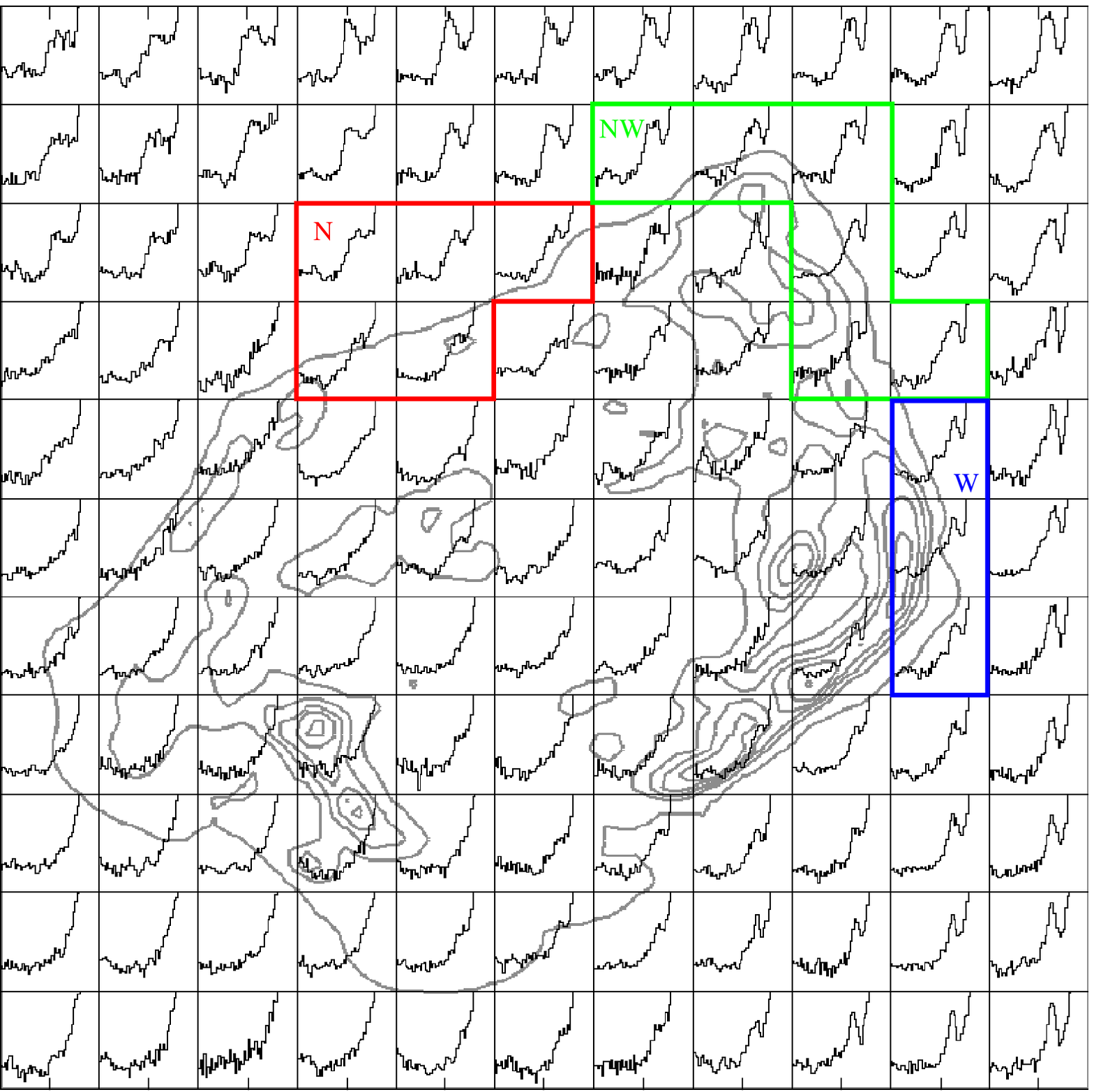,height=0.9\textwidth,angle=0, clip=}
}
\caption{
Grid of \twCO\ ($J$=1--0) spectra restricted to the velocity range
20--40~km~$\ps$, superposed on the VLA 1.4~GHz contours of 3C~397 remnant.
Three regions (``N", ``NW", and ``W") are defined for CO-spectrum
extraction, of which the $32\km\ps$ \twCO\ ($J=1$--0) line profiles
show broad blue wings (see Figure~\ref{f:broaden}).
}
\label{f:grid}
\end{figure*}

\subsection{Kinematic Evidence for Interaction} \label{sec:kinematic}

\begin{figure*}
\centerline{
\psfig{figure=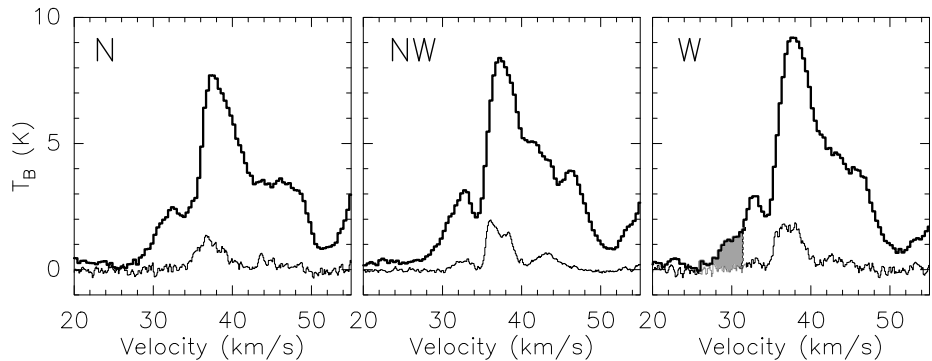,height=0.26\textwidth,angle=0, clip=}
}
\caption{
Averaged CO line profiles of the three regions marked in Figure~\ref{f:grid}.
The \twCO\ line profiles are plotted with thick lines and the
\thCO\ line profiles with thin lines.
The shadowed patch indicates the blueward broadened wing of
the $\sim32\km\ps$ \twCO\ component of region ``W".
}
\label{f:broaden}
\end{figure*}

A grid of \twCO\ ($J$=1-0) spectra is produced 
focusing on the $32\km\ps$ component (Figure~\ref{f:grid}).
We inspected the local spectra along the edges of the remnant
and found that the blue (left) wings ($\sim28$--$31\km\ps$) of the
\twCO\ line profiles of the $32\km\ps$ component in the
northern, northwestern, and westmost edges (see regions
``N'', ``NW'', and ``W'' marked in Figure~\ref{f:grid})
appear to be broadened, as shown in Figure~\ref{f:broaden}.
The blue wings of other positions along the SNR boundary
are not seen to be similarly broadened.
In other cases of SNRs such as IC443 (White et al.\ 1987),
W28 (Arikawa et al.\ 1999),
G347.3$-$0.5 (Moriguchi et al.\ 2003),
Kes~69 (Zhou et al.\ 2009), and Kes~75 (Su et al.\ 2009),
the \twCO\ line broadenings show that the surrounding molecular gas
suffers a perturbation and are regarded as strong kinematic
evidence for the SNR shock-MC interaction.
The broad line profiles seen in 3C~397 are possibly evidence of
such an interaction.
We cannot conclude whether the red (right) wings of the $32\km\ps$ lines
in the three regions
are broad or not, because they are overlapped by the
left wings of the strong $38\km\ps$ lines, which makes it
difficult to determine
the broadness of the red sides of the $32\km\ps$ lines.

In the regions ``N'' and ``NW'' of 3C~397, the broad left wings
may include the contribution from real line broadening due to shock
perturbation.
In the northern edge of the SNR, there are some X-ray structures
that seem to be interacting with the cloud,
such as a hat-like Si knot and an S shell
(according to the X-ray equivalent line width study in Jiang \& Chen 2009).
However, we also note that in a wide northern and northeastern area
in the FOV (see Figure~\ref{f:grid}),
some \twCO\ emission peaks at $30\km\ps$ and thus
may more or less contaminate the broad blue wings
of line profiles of regions ``N'' and ``NW''.

In the westmost region ``W'', there is a unique plateau in the broad blue
wing. No significant \twCO\ emission at $30\km\ps$ is seen near this region.
Therefore the broad blue wing can be safely ascribed to Doppler broadening
of the $32\km\ps$ line, which provides solid kinematic evidence
of interaction between SNR~3C~397 and the adjacent $32\km\ps$ MC.
It is noteworthy that this westmost line-broadening region is essentially
located at a right-angle corner of the rectangular-shaped SNR,
which is close to the western radio and X-ray brightness peaks and
seems to be coincident with the western end of
the Fe-rich ejecta along the ``diagonal"
as revealed in the X-ray equivalent width maps (Jiang \& Chen 2009).
Thus, as a possibility, the molecular gas in this region
might be impacted by the Fe-rich ejecta.

The fitted and derived parameters for the $32\km\ps$ molecular
component in region ``W" are summarized in Table~1.
Here we have applied two methods to estimate the H$_2$ column density and
the molecular mass. In the first method, the X-factor
$N ({\rm H}_2)/W($\twCO$)\approx 1.8\times10^{20}\cm^{-2}{\rm K}^{-1}\km^{-1}$s
(Dame et al.\ 2001) is used.  In the second one,
local thermodynamic equilibrium for the gas and
optically thick condition for the $^{13}$CO ($J$=1--0) line are
assumed, and another conversion relation
$N$(${\rm H}_2)\approx7\E{5}N(^{13}$CO) (Frerking et al.\ 1982) is used.


In view of the morphological correspondence, the broadened line profile,
and the HI~SA comparison,
we suggest that SNR 3C~397 is associated and interacting
with the $32\km\ps$ MC,
which is at a distance of $\sim10.3\kpc$,
and parameterize the distance to the SNR as $d=10.3\du\kpc$.
The distance values used in the previous X-ray studies (e.g., S05)
are numerically very similar to the distance determined here.

\section{DISCUSSION}
\subsection{Mean Molecular Density}\label{sec:density}

\begin{figure}
{\hfil
\psfig{figure=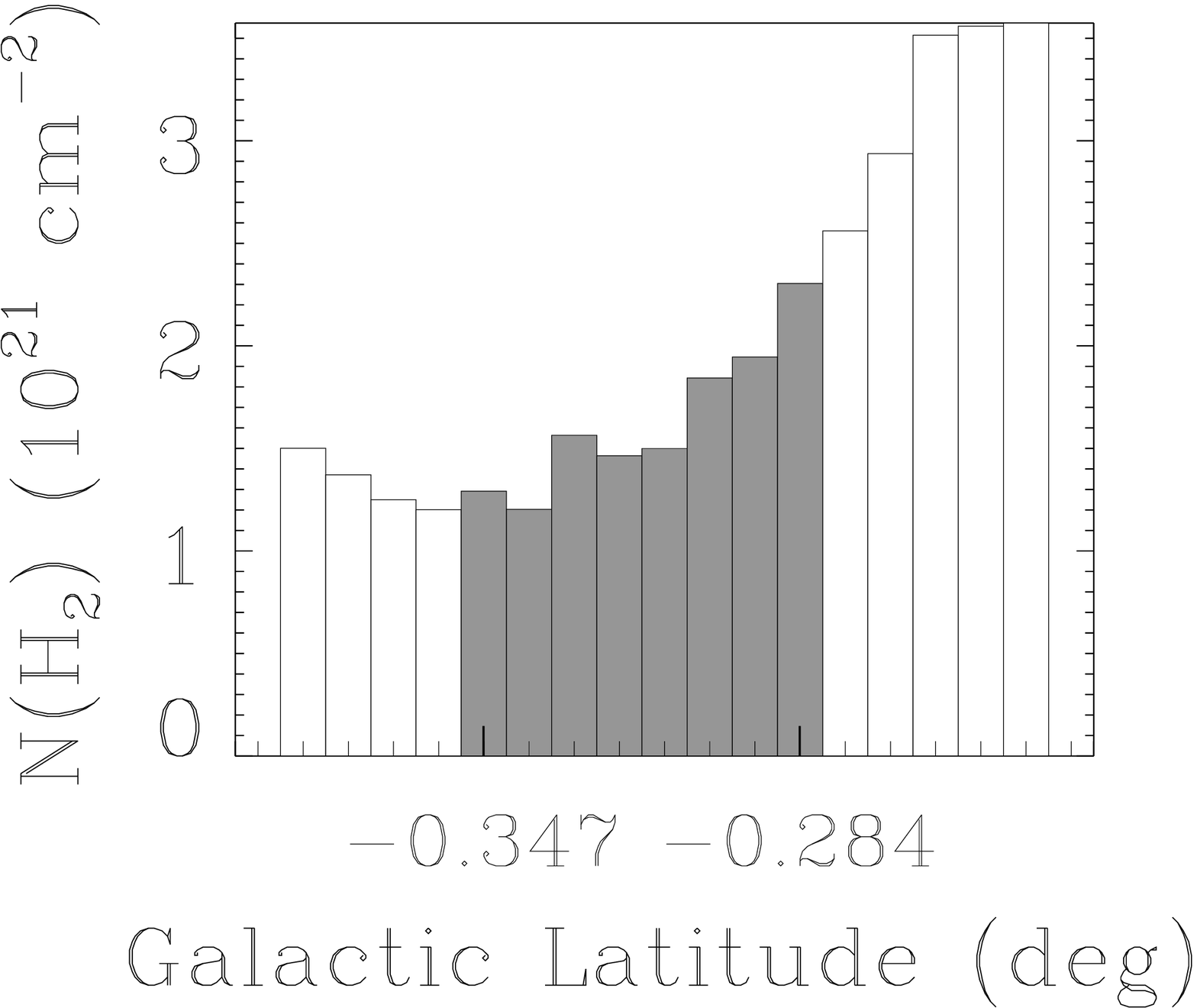,height=0.19\textwidth,angle=0, clip=}
\hspace{1mm}
\psfig{figure=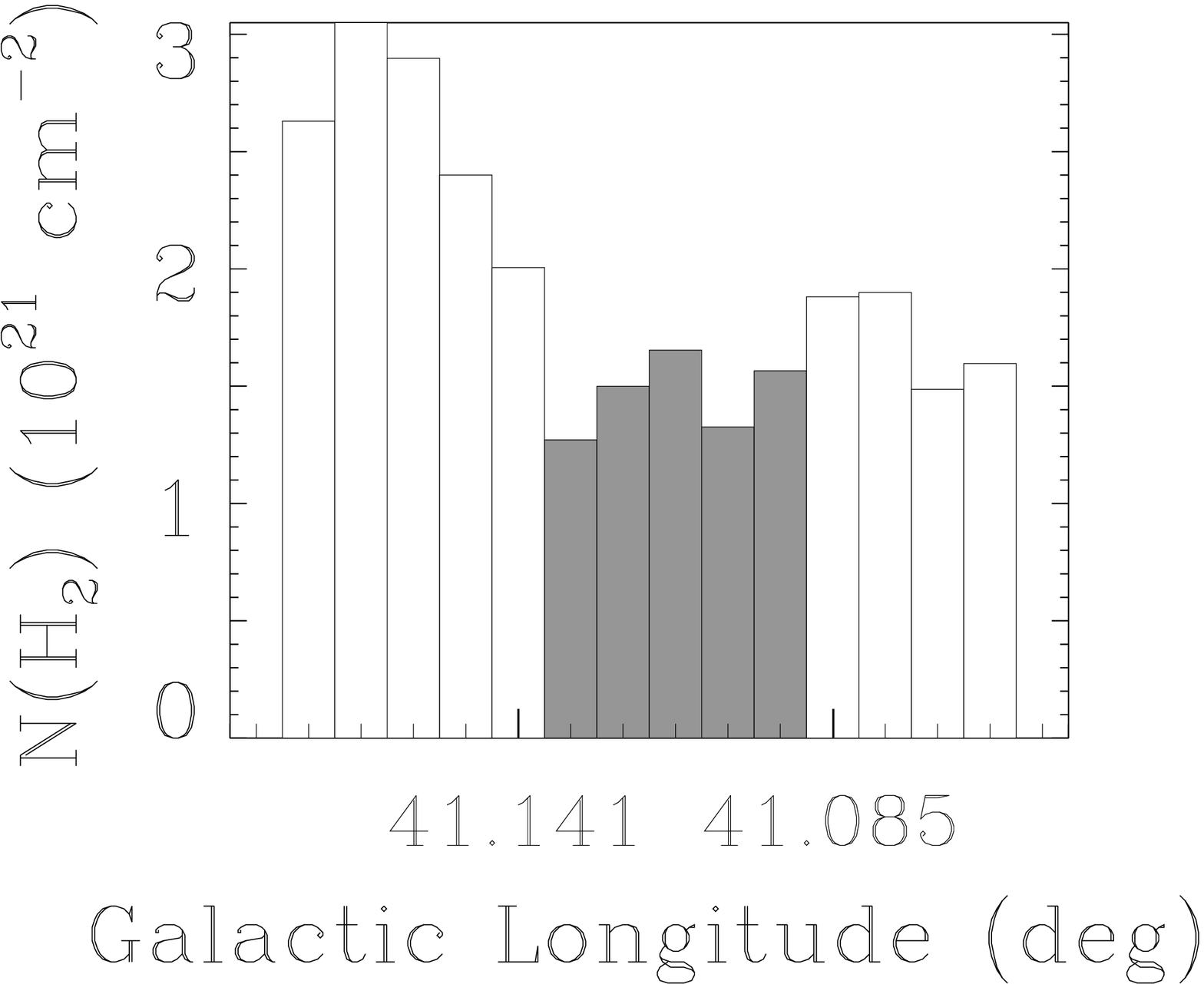,height=0.19\textwidth,angle=0, clip=}
\hfil}
\caption{
The $N$(H$_2$) distributions along the
Galactic latitude (left panel) and longitude (right panel) lines
projected across the remnant (shown in Fig.~\ref{f:vmap}).
The gray bars indicate the extent of the remnant.
}
\label{f:column}
\end{figure}

The H$_2$ column density distributions of the
$V_{\rm LSR}\sim32\km\ps$ component along the Galactic latitude
and longitude lines crossing the remnant center
are plotted in Figure~\ref{f:column}
(using the X-factor method).
The H$_2$ column is shown to increase along the latitude towards
the Galactic plane and has a depression
within the SNR's extent
($\Delta \NHH\sim0.3$--$1\E{21}\cm^{-2}$)
along the longitude, corresponding to the molecular cavity.
This column depression implies a mean
density $\nHH\sim10$--$30\du^{-1}\cm^{-3}$
for the molecular gas which was originally in the cavity.
It can be roughly regarded as the mean density of the environs.
Here we have assumed the line-of-sight length of the cavity as
$10\du$~pc, according to the $3.2'\times4.7'$ angular size of the
rectangular-shaped SNR which converts to a physical extent of
$\sim9\du\parsec\times14\du\parsec$.
For comparison, the $32\km\ps$ molecular component in the FOV
has an average H$_{2}$ column density of $\sim2.5\E{21}\cm^{-2}$
(estimated from the first method) or $\sim1.1\E{21}\cm^{-2}$
(estimated from the second method),
with the mass
amounting to $3.0\E{4}\Msun$ or $1.3\E{4}\Msun$, respectively.

\begin{center}
\begin{deluxetable*}{lcccc}
\tabletypesize{\small}
\tablecaption{Fitted and Derived Parameters for the MCs around 32 km $\ps$ in Region ``W"$^{\rm a}$}
\tablewidth{0pt}
\tablehead{
 \multicolumn{5}{c}{Gaussian Components} \\ \hline
\colhead{Line} & \colhead{Center (km $\ps$)} & \colhead{FWHM (km $\ps$)} &
\colhead{$T_{\rm peak}$ (K)} & \colhead{$W$ (K km $\ps$)}
}
\startdata
\twCO ($J$=1--0) & 32.7 & 2.5 & 2.9 & 7.7\\
\thCO ($J$=1--0) & 32.8 & 1.9 & 0.4 & 0.8 \\ \hline
\\
\multicolumn{5}{c}{Molecular Gas Parameters} \\ \hline
\colhead{} & \colhead{$N$ (H$_2$) $(10^{20}\cm^{-2})$} &
\colhead{$M (\Msun)$} &
\colhead{$T_{\rm ex}$ (K)$^b$} & \colhead{$\tau$ (\thCO)$^c$} \\ \hline
Gaussian components: & 13.9/11.1$^d$ & $197\du^2$/$157\du^2$$^d$ & 8.8 
& 0.17 \\
Residual part$^e$: & 5.2 & $74\du^2$ \\
Total (Gaussian+Residual): & 19.1 & $271\du^2$ 
\enddata
\tablecomments{}
 \tablenotetext{a}{\ The region is marked in Figure~\ref{f:grid}.}
 \tablenotetext{b}{\ The excitation temperature calculated from the maximum
 \twCO ($J$=1--0) emission point of the $32\km\ \ps$ component in the FOV.}
 \tablenotetext{c}{\ The optical depth of the \thCO ($J$=1--0).}
 \tablenotetext{d}{\ Two methods are applied for the estimates (see text in \S~\ref{sec:kinematic}).}
 \tablenotetext{e}{\ Determined by subtracting the Gaussian components
    centered at 32.7~km~$\ps$ from the \twCO\ emission in the velocity
    interval 27--34~km~$\ps$ and applying the X-factor method.}
\end{deluxetable*}\label{tab:west}
\end{center}

The column variation of the $32\km\ps$ cloud
along the latitude is
as large as $\Delta\NHH\sim 1\times 10^{21}\cm^{-2}$,
but it can not alone account for the variation of
intervening hydrogen column density ($\Delta\NH\sim 1.1\E{22} \cm^{-2}$)
from west to east
inferred from the previous \Chandra\ X-ray analysis (S05).
On the other hand, the $38\km\ps$ MC has a
column density $N($H$_2$)$\sim0.6\times10^{22}\cm^{-2}$
(using the first method) or $0.34\times10^{22}\cm^{-2}$ (using the second method),
which is well consistent with the variation of X-ray absorption.
Therefore, the variation in absorption
can basically be explained as a result of the presence of the $38\km\ps$ MC,
which partially covers the western and southern borders of the remnant.

\subsection{Multi-Wavelength Properties}\label{sec:multiwave}

Direct contact of the SNR with the MC can help us understand
the other properties of the remnant.
The pronounced $24\mu$m shell, correlating with the radio shell,
follows the molecular cavity wall rather well.
This mid-IR emission may be primarily due to the dust
collisionally heated by electrons and ions in the outward moving shock
wave as argued for the cases of e.g., N132D (Tappe et al.\ 2006) and
3C~391 (Su \& Chen 2008) where the blast wave has swept up dense material
in those molecular environments.
We refer the reader to Tappe et al.\ (in preparation for ApJ)
for a detailed study of the IR properties of 3C~397.

S05 pointed out that 3C~397 bears some interesting similarities to
the thermal composite (or mixed-morphology) SNR 3C~391 in radio
and X-rays.
The association of 3C~397 with the $32\km\ps$ cloud
demonstrates more similarities between the two SNRs.
Both of them are observed to be located at the edge of a MC,
align in a density gradient
from SE to NW, and represent a ``breakout'' morphology in the east,
resulting in bright radio shells in the NW and lower surface brightness
in the SE.

The association of SNR 3C~397 with the MC
enables us to explain the large volume
emission measure of the low-temperature component of the X-ray emitting
gas (Chen et al.\ 1999; Safi-Harb et al.\ 2000; S05). In fact,
the molecular gas density ($\sim10$--$30\du^{-1}\cm^{-3}$)
is similar to the ambient gas density inferred from the X-ray
spectral analysis (Chen et al.\ 1999; Safi-Harb et al.\ 2000; S05).
It has been suggested based on detailed \Chandra\ X-ray spectroscopic
analysis that the X-ray emission arises from low-ionization-timescale
shock-heated ejecta mixed with shocked ambient/circumstellar material
within a relatively young age of the remnant $\sim5.3$~kyr (S05).
If the molecular cavity was excavated by the supernova blast wave, the ejecta
may be expected to have been fully mixed with the large-mass
($\sim1\E{3}\du[n({\rm H}_2)/20\cm^{-3}]\Msun$)
of swept-up dense gas, so that it would be difficult to spectrally
resolve metal-rich ejecta.
If the cavity was sculpted chiefly by the progenitor star
due to its powerful stellar wind and ionizing radiation,
the ejecta would travel in the low-density region
before impacting the dense molecular gas.
Thus the interaction of the ejecta with the dense ambient gas
should be a recent event and hence the ionization timescale of
the ejecta is naturally low.
This wind cavity scenario seems to be in favor of the suggestion
of reflected shock (S05) inside the SNR.
The impact with the cavity wall may have hampered the expansion of
the ejecta and, while the outermost ejecta can interact with the
dense cloud, the innermost ejecta can be compressed and heated by the
reflected shock (as happens, for example, in W49B [Miceli et al.\ 2006],
in the Cygnus Loop[Levenson et al.\ 2005], and in Kes~27 [Chen et al.\ 2007]).
A detailed investigation of the X-ray properties of the remnant
using \XMM\ will be presented elsewhere (Safi-Harb et al.,
in preparation for ApJ).

\subsection{Multi-Phase Molecular Environment} \label{sec:multiphase}

For the westmost line-broadened region,
we adopt the observed total mass $\sim270\du^2 M_{\odot}$
and the Gaussian-subtracted mass $\sim74\du^2 M_{\odot}$
(Table~1) as the upper and lower limits of the mass of the
disturbed molecular gas, respectively.
Both estimates are larger than the mass,
18--$54\du^2\Msun$, of the molecular gas
that could be swept up from the geometric center to this position
(here we adopt 1/4th the volume of the cone subtended by the western
 boundary since region ``W'' shown in Figure~6 spans approximately half the
 western boundary and since only the one-sided line
 broadening is taken into account).
On the other hand, we have mentioned that the original molecular gas
in the cavity volume can roughly account for the low-temperature
component of the X-ray emitting gas.
Therefore, the disturbed gas in the line-broadened region
is not swept-up gas but preexists there.
The disturbed gas moves at a bulk velocity $v_m$ around $7\km\ps$,
which is implied by the blueward line broadening that reflects the
velocity component along the line of sight.
This motion can be naturally explained to be
driven by the transmitted shock in dense clouds.
The gas may thus be assumed to be in a crude pressure balance
with the proximate hot gas, which will be represented
in the following quantitative analysis
by the ``western lobe'' described in S05.
For the ``western lobe'', the assumption of pressure balance between
the low-temperature component of hot gas ($kT_l\sim0.2\keV$,
with volume emission measure $4.2\times10^{60}\du^2\cm^{-3}$)
and the high-temperature component
($kT_h\sim1.4\keV$, with volume emission measure
$6.2\times10^{57}\du^2\cm^{-3}$) (S05)
yields values of filling factor and atomic density
($f_l\sim0.94$, $n_l\sim66\du^{-1/2}\cm^{-3}$)
for the low temperature component and ($f_h\sim0.06$,
$n_h\sim10\du^{-1/2}\cm^{-3}$)
for the high-temperature component.
(The small filling factor of the high temperature component
seems to be consistent with the ejecta.)
Here we have assumed
$f_l+f_h=1$ (Rho \& Borkowski 2002), while in reality
$f_l+f_h$ may be somewhat smaller than unity.
However, because the hot gas density is proportional to
the inverse of the square root of the filling factor,
the density estimate is not sensitive to the factor.
Hence the above density estimates are lower limits,
but are still a good approximation.
We thus find the density of the disturbed molecular gas
$\nHH\sim2\E{4}\du^{-1/2}\cm^{-3}$
based on the pressure balance between the
cloud shock and the X-ray emitting hot gas
(Zel'dovich \& Raizer 1967; McKee \& Cowie 1975):
$1.4\times2\nHH m_H v_m^2\sim2.3n_l kT_l\,\, (\sim2.3n_h kT_h)$.
The derived high $\nHH$ value implies that
the disturbed molecular gas is in very dense clumps.
The coexistence of moderate-density ($\la100\cm^{-3}$)
molecular component and dense clumps ($\sim10^{4}\cm^{-3}$)
illustrates a complicated multi-phase molecular environment
in the west of SNR 3C~397.
This is not uncommon;
for example, similar multi-phase MCs have also been discovered
in the western region of SNR 3C~391 (Reach \& Rho 2000).

\section{SUMMARY}
We have presented an observation in millimeter CO lines towards the
Galactic SNR 3C~397 which is characterized by an unusual rectangular
morphology.
The SNR is confined in a cavity of molecular gas at an LSR velocity
of $\sim32\km\ps$ and is embedded at the edge of the MC.
The cloud has a column density gradient increasing from SE to NW,
perpendicular to the Galactic plane, and in agreement with
the elongation direction of the remnant.
The $\sim32\km\ps$ \twCO\ line profiles with broad blue wings
are demonstrated along the northern, northwestern, and westmost boundaries
of the SNR; the blue wing of the westmost region may safely be ascribed to
Doppler broadening and hence
shows solid kinematic evidence for the disturbance by the SNR shock.
The systemic velocity, $32 \km\ps$, leads to a determination of
kinematic distance of $\sim10.3\kpc$ to the cloud and SNR 3C~397.
The mean molecular gas density of the SNR environment
is estimated as $\sim10$--$30\du^{-1}\cm^{-3}$, consistent with
the density of the ambient gas
previously inferred from X-ray analyses,
while the disturbed gas density is deduced as $\sim 10^4 \cm^{-3}$,
ascribed to very dense clumps,
implicating a multi-component molecular environment there.
Another MC along the line of sight at around
$38\km\ps$, which was originally suggested to be associated with
the remnant, may be instead located in front of the tangent point,
at a distance of $\sim2.1\kpc$, as implied by the HI~SA.
The variation of X-ray absorption from west to east revealed by S05
can basically be explained as a result of the presence of the
$38\km\ps$ MC, which partially covers the western and southern
borders of the SNR.
We compile a list of Galactic SNRs so far confirmed and suggested to be
in physical contact with adjacent MCs in the appendix.
\\

We thank the anonymous referee for helpful comments.
We are grateful to the staff of Qinghai Radio Observing Station
at Delingha for help during the observation and to Lawrence Rudnick
for providing the VLA data of SNR 3C~397.
Fabrizio Bocchino and Estela Reynoso are appreciated for constructive
help in improving the table of interacting SNRs.
Y.C.\ acknowledges support
from NSFC grants 10725312 and 10673003 and the 973
Program grant 2009CB824800. S.S.H. acknowledges support by
the Natural Sciences and Engineering Research Council of Canada
and the Canada Research Chairs program.
We acknowledge the use of the VGPS data;
the National Radio Astronomy Observatory is a facility of the
National Science Foundation operated under cooperative agreement
by Associated Universities, Inc.
This research made use of NASA's Astrophysics Data System and of
the High-Energy Astrophysics Science Archive Research Center
operated by NASA's Goddard Space Flight Center.

\appendix
\section{Galactic SNRs in Contact with MCs}

Table~2 shows a compilation of Galactic supernova remnants (SNRs)
that are presently known and suggested to be in physical contact with
molecular clouds (MCs).
The evidence for the contact/interaction adopted in the table includes\\
\mbox{\hspace{0.6cm}}
   1. detection of 1720~MHz OH maser within the extent of SNR;\\
\mbox{\hspace{0.6cm}}
   2. presence of molecular (CO, HCO+, CS, etc.) line broadening or
      asymmetric profile (LB); \\
\mbox{\hspace{0.6cm}}
   3. presence of line emission with
      high high-to-low excitation line ratio,
      e.g.\ \twCO\ $J=$2--1/$J=$1--0; \\
\mbox{\hspace{0.6cm}}
   4. detection of near-infrared (NIR) emission (e.g., [FeII] line) or
      vibrational/rotational lines of H$_2$
     [e.g., H$_2$ 1--0 S(1) line (2.12 $\mu$m), H$_2$ 0--0 S(0)-S(7)
     lines] due to shock excitation;\\
\mbox{\hspace{0.6cm}}
   5. specific infrared (IR) colors suggesting molecular shocks,
      e.g.\ Spitzer IRAC 3.6$\mu$m/8$\mu$m, 4.5$\mu$m/8$\mu$m, and 5.8$\mu$m/8$\mu$m
      (Reach et al.\ 2006);\\
\mbox{\hspace{0.6cm}}
   6. morphological agreement (MA) or correspondence of
   molecular features with SNR features
   (e.g.\ arc, shell, interface, etc.).

Condition 1 has been known as a reliable signpost of interacting
SNRs and the combination of condition 6 and one of conditions 2-5 is now
also accepted as convincing evidence for SNR-MC interaction. Condition 6
alone is, however, indicative of probable contact.
%
%
Either of condition 5 and rough spatial coincidence (RC) between
SNR and molecular features is suggestive of possible contact.
The SNRs listed in Table~2 are thus classified into three groups:
34 confirmed ones (``Y''), 11 probable ones (``Y?'') with strong
evidence but not conclusive yet, and 19 possible ones (``?'')
remaining to be determined with further observations.
Note that Tycho SNR is the only known Type~Ia SNR in the table.

Since SNR-MC interaction is an important source of $\gamma$-ray
emission via decay of pions, $\gamma$-ray detections along the line
of sight are also listed in the table.

\begin{center}
\begin{deluxetable}{lllllcl}
\tabletypesize{\scriptsize}
\tablecaption{Galactic SNRs in Physical Contact with MCs}
\tablewidth{0pt}
\tablehead{
\colhead{Name} & \colhead{Other Name} & \colhead{Type$^{\rm a}$} &
\colhead{Evidence$^{\rm b}$} & \colhead{Ref.$^{\rm c}$} & \colhead{Group$^{\rm d}$}&
\colhead{$\gamma$-ray detection$^{\rm e}$(Ref.$^{\rm f}$)}
}
\startdata
G0.0+0.0 & Sgr A East & TC & OH maser, CS MA \& LB, H$_2$ & 1,2,3,4,5 & Y & HESS(67)\\
G1.05-0.1 & Sgr D SNR & S & OH maser & 2,6 & Y & \\
G1.4-0.1 & & S & OH maser & 2,6 & Y & \\
G5.4-1.2 & Milne 56& C? & OH maser & 7 & Y & \\
G5.7-0.0 & & ? & OH maser & 7 & Y & HESS(68) \\
G6.4-0.1 & W28 & TC & OH maser,CO MA \& LB,H$_2$ MA,NIR & 2,8,9,10 & Y & EGRET(69),HESS(68) \\
G8.7-0.1 & W30 & TC & OH maser & 7 & Y & HESS(70) \\
G9.7-0.0 & & S & OH maser & 7 & Y & \\
G16.7+0.1 & & C & OH maser, CO MA & 2,11,12 & Y &  \\
G18.8+0.3 & Kes 67 & S & CO MA \& LB, CO ratio & 13,14 & Y  & \\
G21.8-0.6 & Kes 69 & TC & OH maser,CO MA \& LB,HCO+,H$_2$ & 2,11,15,16 & Y & \\
G29.7-0.3 & Kes 75 & C & CO MA \& LB & 17 & Y &\\
G31.9+0.0 & 3C 391 & TC & OH maser,molecular MA \& LB,H$_2$,NIR & 2,18,19,20 & Y &\\
G32.8-0.1 & Kes 78 & S & OH maser & 21 & Y &\\
G34.7-0.4 & W44 & TC & OH maser, molecular LB, H$_2$ MA, & 2,8,10, & Y & EGRET(69)\\
 & & & NIR, CO ratio & 22 & & \\
G39.2-0.3 & 3C 396 & C & H$_2$ \& NIR MA, CO MA \& LB & 16,23,24 & Y &\\
G41.1-0.3 & 3C 397 & TC & CO MA \& LB & 25 & Y &\\
G49.2-0.7 & W51 & TC & OH maser,CO MA \& LB,HCO+ LB & 2,11,26 & Y & HESS(71),Milagro(72) \\
G54.4-0.3 & HC40 & S & CO MA \& LB, IR MA & 27,28 & Y & \\
G89.0+4.7 & HB21 & TC & CO MA \& LB, CO ratio, H$_2$, NIR & 29,30,31 & Y & \\
G109.1-1.0 & CTB 109 & S & CO MA \& LB & 32 & Y & \\
G189.1+3.0 & IC 443 & TC & OH maser, CO ratio, H$_2$, & 2,8,22,33, & Y & EGRET(69),MAGIC(73)\\
& & & molecular MA \& LB & 34,35 & & Milagro(72),VERITAS(74)\\
 & & & & & & AGILE(75) \\
G304.6+0.1 & Kes 17 & S & H$_2$,IR MA \& colors & 16,28 & Y & \\
G332.4-0.4 & RCW 103 & S & IR MA \& colors,NIR,H$_2$ \& HCO+ MA & 28,36,37 & Y &\\
G337.0-0.1 & CTB 33 & S & OH maser & 18 & Y &\\
G337.8-0.1 & Kes 41 & S & OH maser & 21 & Y &\\
G346.6-0.2 & & S & OH maser, H$_2$, IR colors & 21,16,28 & Y &\\
G347.3-0.5 & & S? & CO MA \& LB & 38 & Y & CANGAROO(76)\\
 & & & & & & HESS(77),Fermi(78)\\
G348.5-0.0 & & S? & OH maser, H$_2$, IR MA & 2,16,28 & Y &\\
G348.5+0.1 & CTB 37A & S & OH maser, CO MA & 2,12,18 & Y & HESS(79)\\
G349.7+0.2 &  & S & OH maser,CO MA \& LB, CO ratio, & 2,18,13, & Y &\\
 & & & H$_2$, IR MA & 16,28 & &\\
G357.7+0.3 & Square Nebula & S & OH maser & 2,6 & Y &\\
G357.7-0.1 & MSH 17-39& TC & OH maser, CO \& H$_2$ MA & 2,18,39 & Y &\\
G359.1-0.5 & & TC & OH maser, CO \& H$_2$ MA & 2,40,41,42 & Y & HESS(80)\\
G33.6+0.1 & Kes 79 & TC & CO MA, HCO+ MA & 43 & Y? &\\
G40.5-0.5 & & S & CO MA & 44 & Y? & Milagro(81),HESS(82)\\
G43.3-0.2 & W49B & TC & H$_2$ MA & 45 & Y? &\\
G54.1+0.3 &  & F? & CO MA & 46 & Y? & \\
G74.0-8.5 & Cygnus Loop & S & CO MA & 47 & Y? & \\
G78.2+2.1 & $\gamma$ Cygni SNR & S & CO MA & 48 & Y? & EGRET(83),Milagro(81)\\
G84.2-0.8 & & S & CO MA & 49,50 & Y? &\\
G120.1+1.4$^{\rm g}$ & Tycho,SN1572 & S & CO MA & 51 & Y? &\\
G132.7+1.3 & HB3 & TC & CO MA & 52 & Y? & EGRET(83)\\
G263.9-3.3 & Vela & C & CO MA & 53 & Y? & CANGAROO(84)\\
 & & & & & & HESS(85)\\
G284.3-1.8 & MSH 10-53 & S & CO MA \& possible LB & 54 & Y? &\\
G11.2-0.3 & & C & IR MA \& colors & 28 & ? & \\
G22.7-0.2 & & S & IR RC & 28 & ? &\\
G23.3-0.3 & W41 & S & CO RC & 55 & ? & HESS(67,70,86),MAGIC(87)\\
G39.7-2.0 & W50, SS433 & ? & CO RC & 56 & ? & \\
G63.7+1.1 & & F & CO RC & 57 & ? & \\
G74.9+1.2 & CTB 87 & F & CO RC & 49,58,59 & ? & \\
G94.0+1.0 & 3C434.1 & S & CO RC & 49 & ? & \\
G106.3+2.7 & & C? & CO RC & 60 & ? & EGRET(83),Milagro(81),\\
 & & & & & & VERITAS(88)\\
G111.7-2.1 & Cas A, 3C461 & S & H$_2$CO absorption,CO RC & 61,62 & ? & HEGRA(89),MAGIC(90),\\
 & & & & & & VERITAS(91),Fermi(92)\\
G160.4+2.8 & HB9 & S & CO RC & 49 & ? &\\
G166.0+4.3 & VRO 42.05.01 & TC & CO RC & 49 & ? &\\
G166.3+2.5 & OV 184 & ? &  CO RC & 49 & ? &\\
G192.8-1.1 & PKS 0607+17 & S & CO RC & 49 & ? &\\
G205.5+0.5 & Monoceros Nebula & S & CO RC & 63 & ? & EGRET(83),HESS (93)\\
G260.4-3.4 & Puppis A & S & CO RC & 64,65 & ? & \\
G290.1-0.8 & MSH 11-61A & TC & CO RC & 66 & ? & \\
G310.8-0.4 & Kes 20A & S & IR RC \& colors & 28 & ? & \\
G311.5-0.3 & & S & IR MA \& colors & 28 & ? & \\
G344.7-0.1 & & C? & IR RC \& colors & 28 & ? & \\
\enddata

 \end{deluxetable}\label{tab:SNRMC}
\end{center}

\clearpage

\begin{center}
\begin{deluxetable}{lllllcl}
\tablewidth{0.96\textwidth}
\tablecaption{Galactic SNRs in Physical Contact with MCs}
\tablecomments{Footnotes in Table~2:}
\tablenotetext{a}{\ Types of SNRs --- S: shell type; C: composite type;
 F: plerion (Crab-like) type;
 TC: thermal composite (Jones et al.\ 1998) (i.e., mixed-morphology,
 Rho \& Petre 1998) type;
 ?: not classified or not sure yet.}

\tablenotetext{b}{\ See the text in Appendix for explanations.}

\tablenotetext{c}{\ References.---(1) Yusef-Zadeh et al.\ 1996;
 (2) Hewitt et al.\ 2008; (3) Serabyn et al. 1992; (4) Yusef-Zadeh et al. 2001;
 (5) Lee et al. 2008; (6) Yusef-Zadeh et al.\ 1999; (7) Hewitt et al.\ 2009a;
 (8) Claussen et al.\ 1997; (9) Arikawa et al.\ 1999; (10) Reach et al.\ 2005;
 (11) Green et al.\ 1997; (12) Reynoso \& Mangum 2000; (13) Dubner et al.\ 2004;
 (14) Tian et al.\ 2007b; (15) Zhou et al.\ 2009; (16) Hewitt et al.\ 2009b;
 (17) Su et al.\ 2009; (18) Frail et al.\ 1996; (19) Reach \& Rho 1999;
 (20) Reach et al.\ 2002; (21) Koralesky et al.\ 1998; (22) Seta et al.\ 1998;
 (23) Lee et al.\ 2009; (24) Su et al.\ to be submitted;
 (25) this work; (26) Koo \& Moon 1997; (27) Junkes et al.\ 1992;
 (28) Reach et al.\ 2006; (29) Koo et al.\ 2001; (30) Byun et al.\ 2006;
 (31) Shinn et al.\ 2009; (32) Sasaki et al.\ 2006; (33) Rosado et al.\ 2007;
 (34) Turner et al.\ 1992; (35) Zhang et al.\ 2009;
 (36) Oliva et al.\ 1999; (37) Paron et al.\ 2006; (38) Moriguchi et al.\ 2005;
 (39) Lazendic et al.\ 2004; (40) Uchida et al.\ 1992; (41) Yusef-Zadeh et al.\ 1995;
 (42) Lazendic et al.\ 2002; (43) Green \& Dewdney 1992; (44) Yang et al.\ 2006;
 (45) Keohane et al.\ 2007; 
 (46) Leahy et al.\ 2008; (47) Scoville et al.\ 1977; (48) Fukui \& Tatematsu 1988;
 (49) Huang \& Thaddeus 1986; (50) Feldt \& Green 1993; (51) Lee et al.\ 2004;
 (52) Routledge et al.\ 1991; (53) Moriguchi et al.\ 2001; (54) Ruiz \& May 1986;
 (55) Leahy \& Tian 2008; (56) Huang et al.\ 1983; (57) Wallace et al.\ 1997;
 (58) Cho et al.\ 1994; (59) Kothes et al.\ 2003; (60) Kothes et al.\ 2001;
 (61) Reynoso \& Goss 2002; (62) Hines et al.\ 2004; (63) Oliver et al.\ 1996;
 (64) Dubner et al., 1988; (65) Paron et al., 2008; (66) Filipovic et al.\ 2005.
}
\tablenotetext{d}{\ Group: classification with ``Y'' as confirmed ones,
 ``Y?'' as probable ones, and "?" as possible ones.}
\tablenotetext{e}{\ $\gamma$-ray detection: any $\gamma$-ray detection towards
 SNRs by various $\gamma$-ray detectors, such as HESS, EGRET, Milagro, MAGIC,
 VERITAS, HEGRA and Fermi.
}
\tablenotetext{f}{\ References.---(67) Aharonian et al.\ 2006a;
 (68) Aharonian et al.\ 2008a; (69) Esposito et al.\ 1996;
 (70) Aharonian et al.\ 2005; (71) Feinstein et al.\ 2008;
 (72) Abdo et al.\ 2009; (73) Albert et al.\ 2007a;
 (74) Acciari et al.\ 2009a; (75) Tavani et al.\ 2010;
 (76) Muraishi et al.\ 2000; (77) Aharonian et al.\ 2004;
 (78) Funk, et al.\ 2009; (79) Aharonian et al.\ 2008c;
 (80) Aharonian et al.\ 2008b; (81) Abdo et al.\ 2007;
 (82) Aharonian et al.\ 2009; (83) Casandjian \& Grenier 2008;
 (84) Katagiri et al.\ 2005; (85) Aharonian et al.\ 2006b;
 (86) Tian et al.\ 2007a; (87) Albert et al.\ 2006;
 (88) Acciari et al.\ 2009b; (89) Aharonian et al.\ 2001;
 (90) Albert et al.\ 2007b; (91) Humensky et al.\ 2009;
 (92) Abdo et al.\ 2010; (93) Aharonian et al.\ 2007.
}
\tablenotetext{g}{\ Remnant of Type~Ia supernova explosion.}
\end{deluxetable}
\end{center}



\label{lastpage}

\end{document}